\documentclass{ws-procs11x85}
\usepackage{multicol}
\usepackage{epsf}

\makeindex
\begin{document}

\title{\bf SATURATION AND HIGH DENSITY QCD} 

\author{ A. H. MUELLER}

\address{Columbia University, Department of Physics, 538 West 120th Street,
New York, NY 10027, USA}

\maketitle\abstract{Recent progress in understanding general properties of high energy scattering near the unitarity limit, where high density gluon components of the wavefunction are dominant, is reviewed. The similarity of the QCD problem and that of reaction-diffusion processes in statistical physics is emphasized.  The energy dependence of the saturation momentum and the status of geometric scaling are discussed.}


\baselineskip=13.07pt
\section{Introduction}

This lecture\footnote{Talk given at ``QCD at Cosmic Energies,'' Erice, 29 August-5 September, 2004.} deals with high density QCD where cross sections are near their unitarity limit (saturation).  This is an area where there has been much new understanding and progress recently.  Indeed, thre have been real breakthroughs in understanding between the time this lecture was originally given and the present time.  The very recent developments will be alluded to, but not developed, somewhat later on.  Conceptually, I believe we are near to achieving a rather complete understanding of how unitarity limits are reached in hard QCD processes at asymptotic energies. 

The lecture starts with a brief discussion of high energy soft scattering where a reasonable phenomenology exists but where there is not yet a good connection to QCD.  Then the general issues of hard scattering at high energies are introduced with a focus on the idea of strong gluon fields and high gluon occupation numbers.  The dipole picture of hard scattering is briefly reviewed and the Balitsky JIMWLK and Kovchegov equations obtained.  Next a picture of the solution of the Kovchegov equation is given and the Y-dependence of the saturation momentum given.  Then the problems with the mean field picture exemplified by the Kovchegov equation are given, and the essential non mean field aspects are outlined.  This leads to an interesting relationship between concepts entering high energy scattering in QCD and similar ideas currently under study concerning reaction-diffusion processes in statistical physics.

\section{High Energy ``soft'' Cross Sections}

Over a wide range of energies one can write, say for proton-proton 
collisions,\cite{Don,Lip}

\begin{equation}
\sigma_{PP}^{tot}\simeq c_1 s^{0.07} + c_2/{\sqrt{s}} .
\end{equation}

\noindent What is surprising about this result is that the form does not change significantly between fixed target energies and the highest energies at the Fermilab collider.  One might have expected to see unitarity limitations more apparent in $\sigma$, however, one has to work harder to see that in fact unitarity limits are being reached even though the rate of growth of the total cross section has not changed.

To see unitarity limits easily it is important to introduce an impact parameter dependence.\cite{Ama}  Then 

\begin{equation}
{d\sigma^{tot}\over d^2b} = 2(1-S(b,s))
\end{equation}

\begin{equation}
{d\sigma^{ine\ell}\over d^2b} = 1-S^2
\end{equation}

\begin{equation}
{d\sigma^{e\ell}\over d^2b} = (1-S)^2
\end{equation}

\noindent where I have assumed that the $S$-matrix is real for this simplified discussion.  Then the $S$-matrix can be obtained from proton-proton elastic scattering as

\begin{equation}
S_{PP}(b,s) = 1-{1\over 4\pi^2} \int d^2\Delta_\perp e^{ib\cdot\Delta_\perp}{\sqrt{{d\sigma_{PP}^{e\ell}\over d^2\Delta_\perp}}}
\end{equation}

\noindent with $S$ schematically illustrated in Fig.1.  At higher and higher energies $S^2,$ the probability of not having an inelastic reaction, becomes very small at larger and larger impact parameters.  When one integrates over impact parameters the form (1) mysteriously emerges which does not show the strong unitarity limits, and the energy independence of $S$ for small  $b,$ exhibted in Fig.1.

\begin{figure}[h]
\centerline{\epsfig{file=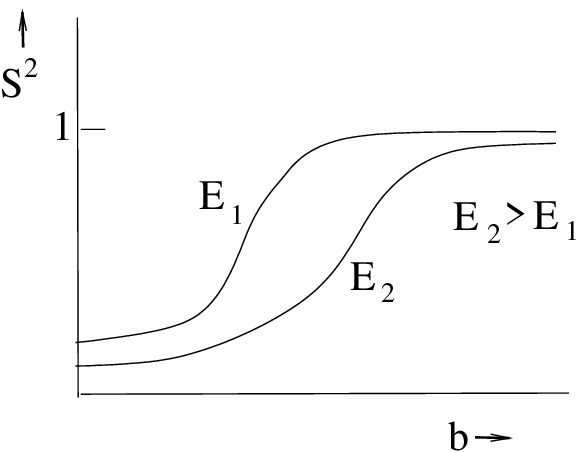}}
\caption{}
\end{figure}

\section{\bf Hard Scattering at High Energies}

Consider the scattering of a small dipole, of transverse size $x_\perp$, on a target at impact parameter  $b.$  The target can be a physical hadron or another dipole.  When  $x_\perp$ is very small\cite{tel}

\begin{equation}
S(b, x_\perp) = 1 - x_\perp^2 {\pi\alpha\over 2N_c}\ {dxG(b, Q^2=4/x_\perp^2)\over d^2b}
\end{equation}

\noindent where the energy (Bjorken-x) dependence has been suppressed.  The $S$-matrix only becomes significantly different from  $1$  when the target gluon distribution becomes of size

\begin{equation}
{dxG\over d^2bd^2k_\perp} \sim {N_c\over \alpha}
\end{equation}

\noindent corresponding to an occupation number for gluons

\begin{equation}
f_g \sim {1\over \alpha N_c}.
\end{equation}

\noindent The scale at which gluon occupation becomes as large  as in (8) and, hence, at which scattering is strong is called the saturation momentum.\cite{bov}  It could be more precisely defined by requiring, say,

\begin{equation}
S(x_\perp = 2/Q_s, b, Y) = 1/2
\end{equation}

\noindent which equation gives $Q_s=Q_s(b,Y)$ with $Y$ the rapidity of the reaction.  In terms of the strength of fields in the target (8) corresponds to $A \sim 1/g$ at the scale of the dipole.

The key questions which arise and which will be considered in much more detail in what follows are: (i)  What is the $Y-$dependence of $Q_s$?  (ii)  What are the $x_\perp$ and $Y-$dependences of $T=1-S$?  We recall that geometric scaling\cite{sto} corresponds to

\begin{equation}
T=T(x_\perp^2Q_s^2,b).
\end{equation}

\section{\bf The Dipole Picture of Hard Scattering;$^7$ the Balitsky-JIMWLK$^{8-10}$ and Kovchegov$^{11}$ Equations}

Consider the scattering of a dipole on a target where the target carries almost all the momentum and is right-moving.  The left-moving dipole has little momentum and serves as a probe of the highly evolved target.  Now let $Y \rightarrow Y + dY$ and put the possible additional evolution into the probe which is a simple object.  The initial dipole probe having a quark line at $\underline{x}_0$ and an antiquark line at $\underline{x}_1$ may have a virtual correction in the interval $dY$, in which case it remains an elementary dipole, or a gluon may now be present at $\underline{x}_2$ thus giving two dipoles.  The two-dipole term consists of a dipole having a quark at $\underline{x}_0$ and an antiquark at $\underline{x}_2$ and another dipole with a quark at $\underline{x}_2$ and an antiquark at $\underline{x}_1.$  This gives the rapidity dependence of the scattering $S-$matrix as

\begin{displaymath}
{dS(x_{01},Y)\over dY}={\alpha N_c\over 2\pi^2}\int{d^2x_2x_{01}^2\over x_{12}^2x_{02}^2}
\end{displaymath}
\begin{equation}
[S^{(2)}(x_{02},x_{12},Y)-S(x_{01},Y)]
\end{equation}

\noindent where the $S^{(2)}-$ term is the $S-$matrix for scattering the two dipole term on the target and the $S-$term is the  virtual term.  The factor ${\alpha N_c\over 2\pi^2}\ {x_{01}^2d^2x_2\over x_{02}^2 x_{12}^2}$ is the element of probability for emitting a gluon, either real or virtual,  and $x_{01} \equiv \vert\underline{x}_0-\underline{x}_1\vert$, etc.  Eq.11 is illustrated in Fig.2.

\vskip 10pt
\begin{figure}[h]
\centerline{\epsfig{file=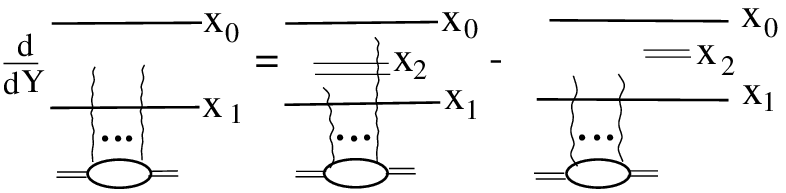}}
\caption{}
\end{figure}
\vskip 10pt

Eq.11 is part of the Balitsky-JIMWLK hierarchy.$^{8-10}$  The difficulty with (11) is that one also needs an equation for $S^{(2)}$ which will couple to a three-dipole scattering term, with the coupling to higher dipole terms never terminating. Kovchegov,\cite{Kov} motivated by scattering off a large nucleus, suggested approximating $S^{(2)}$ by

\begin{equation}
S^{(2)}(x_{02}, x_{12}, Y) = S(x_{02},Y) S (x_{12},Y)
\end{equation}  

\noindent where all the $S-$matrices are to be understood as referring to a given (but unspecified) impact parameter.  Eq.12 is a type of mean field approximation and leads to

$$
{dS(x_{01}, Y)\over dY} = {dN_c\over 2\pi^2}\int{d^2x_2x_{01}^2\over x_{02}^2x_{12}^2}
$$
\begin{equation}
\cdot \ \  [S(x_{02},Y) S(x_{12}, Y) - S(x_{01}, Y)].
\end{equation}

\noindent This equation, the Kovchegov equation, is now a precise equation for the scattering matrix and has a unique solution once the initial condition is fixed.  Introducing $T=1-S$ one may write (13) as

$$
{d\over dY} T(x_{01},Y)= {\alpha N_c\over 2\pi^2} \int {d^2x_2 x_{01}^2\over x_{02}^2x_{12}^2}[T(x_{02}, Y)
$$
\begin{equation}
 + T(x_{12}, Y)-T(x_{01},Y)-T(x_{02}, Y) T(x_{12}, Y)]
\end{equation}

\noindent with the first three terms on the right-hand side of (14) corresponding to the BFKL\cite{Kur,lit} equation and the final term giving the nonlinear behavior which guarantees unitarity for the overall scattering.  While (13) and (14) are not exact equations they are extremely useful as semi-realistic equations having BFKL behavior and obeying unitarity constraints.  This is probably the best one can do in terms of a definite equation for a single function.  The Balitsky-JIMWLK equations keep correlations better than the Kovchegov equation, but even they lack some of the essential ingredients necessary to impose unitarity in a genuinely realistic way.\cite{ncu}

\section{\bf The Solution (and Picture) of Evolution in the Kovchegov Equation}

While the Kovchegov equation is not an exact equation it does lead to interesting results some of which appear to be quite general and others which may be reasonably accurate in a limited energy region.  One of the strongest results coming from the Kovchegov equation concerns the saturation momentun, $Q_s$, introduced in (9).  One finds\cite{bov,ran,Mun,lle}

$$
\ell  n Q_s^2/\mu^2 = {2\alpha N_c\over \pi}{\chi(\lambda_0)\over 1-\lambda_0} Y-{3\ell  n(\alpha  Y)\over 2(1-\lambda_0)}
$$
\begin{equation}
 \ \ \ \ \ + {1\over 1-\lambda_0} \ell n[\alpha^2\ell n 1/\alpha^2]
\end{equation}

\begin{equation}
\ell n Q_s^2/\wedge^2 = {\sqrt{{4\alpha  N_c\over \pi b}{\chi(\lambda_0)\over 1-\lambda_0}Y}} - cY^{1/6}.
\end{equation}

\noindent Eq.(15) refers to a fixed coupling Kovchegov equation while (16) is for running coupling.  The $\mu^2$ in (15) is some (scheme dependent) scale while the  $c$\ in (16) is known in terms of the first zero of an Airy function.  $\chi(\lambda)$ is the BFKL eigenvalue function with $\lambda_0$ satisfying

\begin{equation}
\chi(\lambda_0) = - (1-\lambda_0) \chi^\prime(\lambda_0).
\end{equation}

The first terms in (15) and (16), those involving the ratio ${\chi(\lambda_0)\over 1-\lambda_0},$ were obtained long ago by Gribov, Levin and Ryskin\cite{bov} from BFKL evolution.  The full results were obtained in Ref.17 using a cutoff to simulate the nonlinear terms in the Kovchegov equation.  Munier and Peschanski\cite{Mun} then obtained these full results using the general theory of nonlinear euqations in the FKPP\cite{her,Kol} class.  The FKPP equation is a ``simple'' one-dimensional equation

\begin{equation}
{\partial P(xx,t)\over \partial t} = {\partial^2P\over \partial x^2} + P-P^2
\end{equation}

\noindent which has many of the general features contained in the more complicated Kovchegov equation.  As we shall see a little later on, the leading asymptotic terms in (15) and (16) are correct although the exact results are different in the subleading asymptotic behaviors.

In addition to the energy (rapidity) dependence of the saturation momentum the Kovchegov equation leads to a scaling property of the scattering amplitude which reads\cite{sto}

\begin{equation}
T(x_{01}, Y) \simeq c_1 [Q_s^2(Y)x_{01}^2]^{1-\lambda_0}[\ell  n {1\over Q_s^2x_{01}^2} + c_2]
\end{equation}

\noindent valid when $Q_s^2x_{01}^2 < 1$ but where $Q_s^2x_{01}^2$ is not too small.  The region where the Kovchegov equation gives scaling is quite a large region.\cite{ran}  The experimental data seem to support this scaling.  Unfortunately, scaling is not correct asymptotically in a full treatment in QCD, but it may be true that there is an intermediate energy region where the Kovchegov equation is approximately valid giving (19) in the current HERA energy regime.  This is an important point on which there is little understanding at the moment.

In comparing results from the Kovchegov equation with experimental data one generally needs to use a model of the Golec-Biernat W\"usthoff type.\cite{lec,ura}  How do the results hold up?  Experimentally, ${d\over dY} \ell   n Q_s^2/\mu^2 \simeq 1/4.$  Eq.(15) gives a much larger value for ${d\over dY} \ell   n Q_s^2$ if one takes $\alpha_s\simeq 1/4$, say.  On the other hand, the running coupling answer given in (16) is much closer to experiment.  Triantafyllopoulos\cite{fyl} has done a more complete analysis than that given in (16).  He includes the higher order terms in the BFKL kernel as well as a resummation of collinear effects along the lines of that developed by Ciafaloni, Colferai and Salam.\cite{Cia}  In the end he finds ${d\over dY} \ell  n Q_s^2\simeq 1/4 - {1\over 3}$ in the HERA regime with the (pleasant) surprise being that higher order corrections and resummation do not change the result very much.

\section{\bf Problems with the Mean Field Picture}

While the mean field picture contained in (13) and (14) can be expected to be accurate for scattering on a large nucleus over a limited energy regime these equations cannot in general be expected to be accurate.  Numerical calculations by Salam\cite{Sal} sometime ago have shown that unitarity is reached in a very irregular way in high energy scattering.  At one value of the impact parameter the scattering amlplitude, $T,$ may be small while at another impact parameter  $T$ may be close to its unitarity limit.  A similar phenomenon occurs at a fixed impact parameter when one considers an ensemble of possible wavefunctions.  Some elements of the ensemble will have a a very small  $T$ while other elements of the ensemble may have  $T$ near one.  Thus, an equation like (14), where such fluctuations are neglected,  cannot be expected to lead to correct general results.  What we shall see in the following sections is that the picture of fluctuations in the high energy hard scattering QCD problem (at a fixed impact parameter) closely resembles the fluctuations which are actively being studied by statistical physicists concerned with reaction-diffusion processes.

\section{\bf Beyond the Kovchegov Equation; A First Try}

In trying to understand the domain of validity of the Kovchegov equation the authors of Ref.29 ran into the following dilemma.  Imagine starting from an elementary dipole of size $x_A$ at $y=0.$  Evolve the dipole, using BFKL evolution, to $y=Y$ and ask what is the number density of dipoles of size $x_B$ at $Y.$  This is what one calls $n(x_A, x_B, Y).$  $n$  obeys a completeness relation

\begin{equation}
n(x_A, x_B, Y) = \int {d^2x\over 2\pi  x^2} n(x_A, x, Y/2) n(x, x_B, Y/2),
\end{equation}

\noindent and since

\begin{equation}
T(x_A, x_B, Y) \simeq c \alpha^2 x_B^2n(x_A, x_B, Y)
\end{equation}

\noindent when  $x_B^2 < x_A^2,$ one can approximately rewrite (20) as 

\begin{displaymath}
{1\over x_B^2} T(x_A, x_B, Y) \simeq {1\over 2 c \alpha^2}
\end{displaymath}
\begin{equation} 
\int d\rho ({1\over x^2} T (x_A, x, Y/2))\cdot ({1\over x_B^2} T(x, x_B, Y/2)),
\end{equation}

\noindent where $\rho = - \ell n (x^2\mu^2).$

Suppose we imagine $x_B$ being not too far from the saturation line, as shown in Fig.3.  Then in the approximation of using an absorptive cutoff to simulate the nonlinear term in the Kovchegov equation the evolution described by Eqs.(20) and (22) is simply BFKL evolution with an absorptive boundary at $\rho_1(Y)$ as illustrated in Fig.3.  It is straightforward to see that the region of $\rho$ which dominates the integral in (22) is

\begin{figure}[t]
\centerline{\epsfig{file=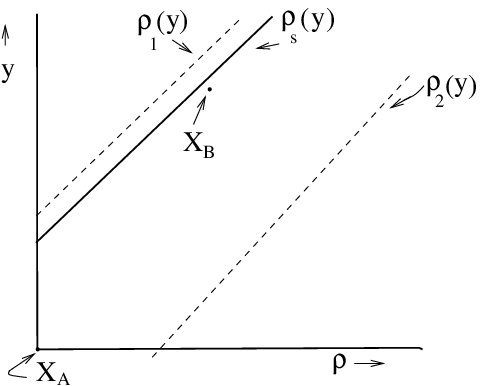}}
\caption{}
\end{figure}

\begin{equation}
\rho-\rho_s(Y/2) \sim {\sqrt{\alpha Y}},
\end{equation}

\noindent the normal diffusion region for BFKL evolution.  But

\begin{equation}
{1\over x^2} T(x_A, x, Y/2) \sim e^{-(1-\lambda_0)(\rho-\rho_s(Y/2))}
\end{equation}

\noindent in this region so that ${1\over x^2} T(x_A, x, Y/2)$ is typically extremely small in the integrand in (22).  But if $x_B$ is not far from the saturation line ${1\over x_B^2} T(x_A, x_B, Y)$ is not small.  This means that ${1\over x_B^2} T(x, x_B, Y/2)$ must be typically very large.  However, this latter fact is a violation of unitarity.  Thus Ref. 29 concluded that while the Kovchegov equation gave unitarity limits on the \underline{overall} evolution of a process it did not guarantee that intermediate stages of the evolution (going from $Y/2$ to $Y$ in the above example) would obey unitarity limits.

The simple procedure used in Ref.29 for ``fixing'' the problem was to limit the $\rho-$integration in (22) so that ${1\over x_B^2} T(x, x_B, Y/2)$ would never be larger than one.  This was done by introducing a second boundary, $\rho_2(Y),$ to the right of which evolution was not allowed.  The resulting problem was then that of simple BFKL evolution in the $\rho-Y$ plane with the evolution being restricted to
lie within the region $\rho_1<\rho <\rho_2$ as illustrated in Fig.3

It is then not a difficult task to find the resulting saturation momentum to be\cite{Sho}

\begin{displaymath}
\rho_s(Y)=\ell  n Q_s^2/\mu^2 = {2\alpha N_c\over \pi} {\chi(\lambda_0)\over 1-\lambda_0}Y
\end{displaymath}
\begin{equation}
\cdot \left[1-{\pi^2\chi^{\prime\prime}(\lambda_0)\over 2(\Delta\rho)^2\chi(\lambda_0)}\right]+ O{\alpha  Y\over (\Delta\rho)^3}
\end{equation}

\noindent where 

\begin{equation}
\Delta\rho = {1\over 1-\lambda_0} \ell   n 1/\alpha^2 + {3\over 1-\lambda_0} \ell  n \ell  n {1\over \alpha^2} + const.
\end{equation}

\noindent Perhaps the bigger surprise is that geometric scaling is lost in the above approach with

\begin{equation}
T(\rho,Y) = T ({\rho-\rho_s(Y)\over \alpha Y/(\Delta\rho)^3})
\end{equation}

\noindent emerging rather than

\begin{equation}
T = T (\rho-\rho_s(Y))
\end{equation}

\noindent which follows from the Kovchegov equation.  As we shall shortly see (25) is expected to be an exact asymptotic result while (27) represents a shortcoming of the ``mean field'' approach of Ref.29.

\section{\bf Interpreting the Second Boundary; An Event by Event Picture of Evaluation} 

Now, following Ref.30 and inspired by work on reaction-diffusion processes in statistical physics,\cite{rid} we view evolution in a different way than we have presented it up to now.  To have a specific example in mind consider the evolution of a ``target'' dipole which is elementary and of size $x_A$ at $Y=0.$  At rapidity $Y$ we probe this evolved diple with an elementary dipole of size $x_B$ with the resulting scattering amplitude $T(x_A, x_B, Y)$ emerging.  $T(x_A, x_B,Y)$ is an average quantity because there are many possible states of the target, starting from an elementary dipole at $y=0,$ which contribute to $T(x_A, x_B, Y).$  For example, in the procedure used by Salam\cite{Sal} to generate evolution of dipole wavefunctions using a Monte Carlo method one must sum over all the events generated at $Y$ to obtain $T(x_A, x_B,Y).$  Now we wish to view evolution of a single event.  For dipole evolution this evolution is governed by the same kernel which appears in (14), but now the kernel gives the probability that a dipole of size $x_{01}$ split into two dipoles of sizes $x_{02}$ and $x_{21}$.  This is a stochastic process since what happens in a given ``time'' interval $dy$ is not certain, but the various outcomes (splittings or no splitting) are given various probabilities, determined by the kernel of Eq.14.  Now denote by $\hat{T} (x_A, x_B,Y)$ the scattering amplitude for a dipole of size $x_B$ scattering  on a particular component of the evolved state starting from $x_A$ at  $y=0.$  We call this a single event and note that

\begin{displaymath}
<\hat{T} (x_A, x_B, Y)> = T(x_A, x_B, Y).
\end{displaymath}

\noindent New view $\hat{T}(x_A, x_B, Y)$ as a function of $\rho = - \ell   n(\mu^2x_B^2).$  That is, for a particular evolution of the target starting from $(x_A)$ we make probes using projectile dipoles of various sizes, $x_B.$  The picture is as illustrated in Fig.4.  The picture is that of a travelling wave with velocity equal to ${d\rho_s(y)\over dy}.$  Since

\begin{equation}
\hat{T} (\rho, y)\simeq c \alpha^2 n (x_A, x_B,y),
\end{equation}

\begin{figure}
\centerline{\epsfig{file=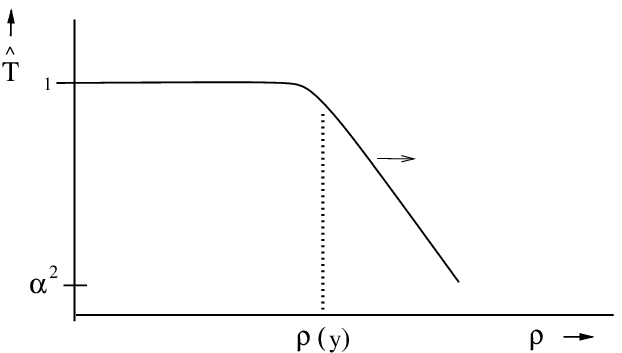}}
\caption{}
\end{figure}

\noindent and since  $n$  is discrete, the function $\hat {T}$ decreases rapidly to zero for $\rho$ greater than $\rho_2$ where 
$\hat {T} (\rho_2,y) = \alpha^2.$  This intereprets the second boundary, discussed in the last section, as a cutoff reflecting the fact that for a given event the gluon occupation number cannot be less than one for any given ``size'' gluon.  For $\rho_s(y) < \rho < \rho_2$ the evolution of $\hat{T}$ obeys the BFKL equation since gluon occupation is high and the stochastic part of the evolution is small.  The approximation of Ref.29 is to neglect the stochastic part of the evolution completely and use BFKL evolution between the boundaries where $\hat{T}$ takes on values near one and near $\alpha^2.$

What one gets out of viewing the process in an event by event way is the understanding that BFKL evolution is correct for the evolution of a single event away from $\hat {T}  =1$ and $\hat {T}  =\alpha^2.$  This gives one the understanding that the (25) should be an exact (asymptotic) result in QCD since the velocity for all travelling waves (events) is the same.

What about the stochastic part of the evolution, the randomness inherent in the dipole splitting when $\hat{T}\simeq \alpha^2$?  The stochasticity is reflected in the fact that different events do not have exactly the same values of $\rho_s,$ as illustrated in Fig.5.  The idea is that fluctuations in the tail of the distribution can lead, after their effect diffuses into the main part of the travelling wave, to fluctuations in the saturation momentum from event to event.  The strength of these fluctuations is given by

\begin{equation}
\sigma^2=<\rho_s^2> - <\rho_s>^2.
\end{equation}

\begin{figure}[t]
\centerline{\epsfig{file=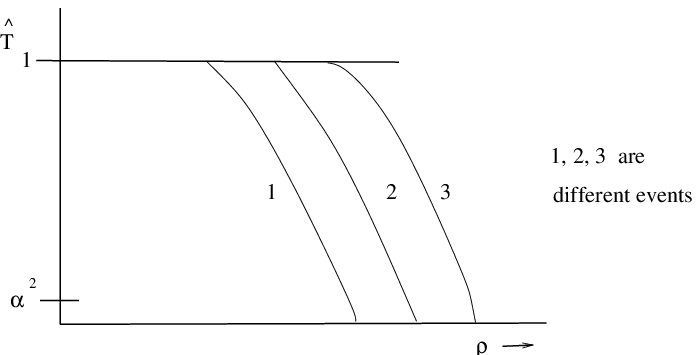}}
\caption{}
\end{figure}

\noindent There have been elaborate numerical studies of $\sigma^2$ in statistical models.  The conclusion is that\cite{rid,oro}

\begin{equation}
\sigma = const.\left[{\alpha Y\over (\Delta \rho)^3}\right]^{1/2}
\end{equation}

\noindent although, as yet, there is no analytical argument leading to this result.  We presume that (31) also holds true for QCD evolution, at a definite impact parameter, and this naturally gives\cite{anc}

\begin{equation}
T(\rho, Y) = T\left({\rho-\rho_s(Y)\over {\sqrt{\alpha/(\Delta\rho)^3}}}\right)
\end{equation}

\noindent where the $\rho_s(Y)$ in (32) is the average, $<\rho_s(Y)>$ of the saturation momentum coming from the various individual events.  Eq.32 differs from (27), where stochastic effects were ignored, and it also represents a violation of geometric scaling.

The studies of statistical systems leading to (25) and (31) also suggest that these limits are reached rather slowly.\cite{rid,oro}  Thus, while (25), (31) and (32) are likely exact results in QCD, although (25) needs some straightforward modifications to take running coupling effects into account, it is not yet clear that these effects can be seen at current energies.  It is a very important task to better understand subasymptotic terms, both to connect with experiment and to see when asymptotia can be expected to arrive. Up to recently it has been thought that the Balitsky-JIMWLK equations are the correct equations for following amplitudes from weak to strong scattering in QCD as energy increases, and there has been an interesting numerical study\cite{Rum} of the Balitsky-JIMWLK equations.  However, it now appears that the Balitsky-JIMWLK equations\cite{sky,Ian} miss essential parts of the dynamics necessary to correctly impose unitarity on BFKL evolution.\cite{ncu}  It will be interesting to see if a new equation emerges which corrects the deficiencies of Balitsky-JIMWLK and yet allow accurate numerical simulations so  that the  approach to asymptopia can be effectively studied.

\section*{Acknowledgments}
This work is supported in part by the Department of Energy.


\end{document}